\documentclass[twocolumn,showpacs,preprintnumbers,amsmath,amssymb,aps]{revtex4}

\usepackage{graphicx}
\usepackage{dcolumn}
\usepackage{bm}

\begin{document}


\title{Unified Models of Inflation and Quintessence}

\author{Andro Gonzalez$^{1,3}$, Tonatiuh Matos$^2$ and Israel Quiros$^{2,3}$}

\email{tmatos@fis.cinvestav.mx;israelquiros@yahoo.com.mx}

\affiliation{$^1$ Universidad de Ciego de Avila. Cuba\\$^2$
Departamento de F{\'\i}sica, Centro de Investigaci\'on y de
Estudios Avanzados del IPN, A.P. 14-740, 07000 M\'exico D.F.,
M\'exico\\$^3$ Universidad Central de Las Villas. Santa Clara CP
54830. Cuba}

\date{\today}

\begin{abstract}
We apply an extended version of the method developed in reference
\cite{chimento}, to derive exact cosmological (flat)
Friedmann-Robertson-Walker solutions in RS2 brane models with a
perfect fluid of ordinary matter plus a scalar field fluid trapped
on the brane. We found new exact solutions, that can serve to
unify inflation and quintessence in a common theoretical
framework.
\end{abstract}


\maketitle

\section{Introduction}

Recent observations from the Wilkinson Microwave Anisotropy Probe
(WMAP) \cite{wmap}, offer strong supporting evidence in favor of
the "Inflationary Paradigm".\footnotemark\footnotetext{As
predicted theoretically within the framework of the conventional
inflationary model of the early universe, these observations
signal at a gaussian, adiabatic and nearly scale-invariant form of
the cosmic microwave background (CMB) power spectrum and at zero
spatial curvature (spatially flat universe).} In the most simple
models in this kind, the energy density of the universe is
dominated by the potential energy of a single (inflaton) scalar
field that slowly rolls down its self-interaction potential
\cite{inflation}. Restrictions imposed upon the kinds of
potentials that can lead to realistic inflationary scenarios, are
dictated by the slow-roll approximation, and result in that, only
sufficiently flat potentials can drive inflation. In order for the
potential to be sufficiently flat, these conventional inflationary
models should be fine-tuned. This simple picture of the early-time
cosmic evolution can be drastically changed if one considers
models of inflation inspired in "Unified Theories", like the Super
String or M-theory. One of the most appealing models of this kind
is the Randall-Sundrum brane world model of type 2 (RS2)
\cite{randall}. In this model a single co-dimension 1 brane with
positive tension, is embedded in a five dimensional AdS (bulk)
space-time, which is infinite in the direction perpendicular to
the brane. In general the standard model (SM) matter degrees of
freedom are confined to the brane, meanwhile, gravitation can
propagate in the bulk. However, in the low-energy limit, due to
the curvature of the bulk, the graviton is confined to the brane,
and standard (four dimensional) general relativity (GR) laws are
recovered.

RS2 braneworld models have an appreciable impact on early universe
cosmology, in particular, for the inflationary paradigm. In
effect, a distinctive feature of cosmology with a scalar field
confined to a RS2 brane, is that the expansion rate of the
universe differs at high energy from that predicted by standard
GR. Actually, due to a term that is quadratic in the energy
density, the friction acting on the scalar field is enhanced. This
means that, in RS2 braneworld cosmology, inflation is possible for
a wider region of parameter space than in standard cosmology
\cite{hawkins}.\footnotemark\footnotetext{This is true for
braneworld cosmology in general.} Even potentials that are not
sufficiently flat from the point of view of the conventional
inflationary paradigm, can produce successful inflation. At
sufficiently low energies (much less than the brane tension), the
standard cosmic behaviour is recovered prior to primordial
nucleosynthesis scale ($T\sim 1 MeV$) and a natural exit from
inflation ensues as the field accelerates down its potential
\cite{lidsey}. In this scenario reheating arises naturally even
for potentials without a global minimum and radiation is created
through gravitational particle production \cite{ford} and/or
through curvaton reheating \cite{feng}. This last ingredient
improves the brane "steep" inflationary picture \cite{urena}.
Other mechanisms as preheating, for instance, have been also
explored \cite{sami}.

Another interesting feature of this scenario, is that the inflaton
does not necessarily need to decay and it may survive through the
present epoch in the cosmic evolution. Therefore, it may play also
the role of the quintessence field, that is a necessary ingredient
to explain the current acceleration in the expansion of the
universe. Such a unified theoretical framework for the description
of both inflaton and quintessence with the help of just one single
scalar field, has been the target of some works (see for instance
references
\cite{lidsey,majumdar,sss,kdimopoulos}).\footnotemark\footnotetext{These
include models of quintessential inflation without branes as in
\cite{peebles}.} However, in general, it is very difficult to
solve the system of Einstein's differential equations that model a
RS2 brane with Friedmann-Robertson-Walker (FRW) metric on it, even
if only a scalar field matter degree is confined to the brane and
the slow-roll assumptions are invoked \cite{hawkins}. In
consequence, it is of interest to develop generating techniques
for deriving exact solutions to Einstein's field equations on the
brane for the aforementioned problem \cite{hawkins}, or to use
well known techniques that have been used to generate solutions in
standard GR, but not in a brane context \cite{chimento,tmatos}. In
Ref.\cite{hawkins}, for instance, the authors develop algorithms
for generating a class of exact braneworld cosmologies, where a
self-interacting scalar field $\phi$ is confined to a RS2
brane.\footnotemark\footnotetext{In \cite{hawkins} the scalar
field $\phi$ is the only matter source.} They found a number of
exact solutions and, in the perfect fluid model the resulting
potential is identified. It behaves like
$V(\phi)\sim\sinh^{-2}[\lambda\phi]$. The algorithms developed are
valid only during inflation, where the field is monotonically
rolling down its potential. In Ref.\cite{savchenko}, the authors
study a RS2 braneworld cosmology for a universe filled with a
perfect fluid of ordinary matter and a scalar field with a
power-law potential $V(\phi)\sim\phi^\alpha$. The index $\alpha$
can be either positive or negative. The authors describe scaling
solutions. Radiation dominated and scalar field dominated
solutions to this model were studied in \cite{kmaeda}. The author
studied the region in the parameter space where a realistic
quintessence model is possible.

The aim of the present paper is to generalize the method for
generating exact FRW cosmologies developed in reference
\cite{chimento}, with the purpose to derive exact solutions in a
RS2 braneworld with a two-component perfect fluid, consisting of
an ordinary fluid plus a scalar-field. This method has been used
to generate FRW solutions in standard (four dimensional) GR
\cite{chimento,tmatos}, and here we generalize it for the brane
contexts. We are able to obtain new exact solutions that
generalize other existing braneworld FRW solutions. A salient
feature of these solutions is that they can accommodate unified
models of both early time inflation and late time accelerated
expansion. Since the solutions are quite general, we are led to
conjecture that such a unified pattern of inflation is generic of
RS2 cosmological models.

The paper has been organized in the following way. In section
\ref{setup}, we expose the details of the braneworld model we are
about to investigate and we explain the method we use for
generating exact solutions. In section \ref{soluciones} we give
the "gallery" of solutions we are able to derive and briefly
comment on some of then, making emphasis on their ability to
explain, in a unified framework, both early inflation and late
time "quintessential" inflation. Section \ref{unifi} is dedicated
to integrate the results in previous section within a unified
picture of inflation. In the final section we summarize the main
results we obtained.

We are interested in the region in the parameter space where the
classical solution is still valid, but where quadratic (brane)
corrections could become significant. This means that the brane
tension could be much less than the energy density of the matter
degrees of freedom which, in turn, is much less than $M_{(5)}^4$,
where $M_{(5)}$ represents the five dimensional Planck scale.

\section{The Set Up}\label{setup}

We study a brane cosmology set up based on RS2 braneworld model
\cite{randall}, in which the brane is filled both with a scalar
field fluid and a perfect fluid of "ordinary" matter. i. e., the
total energy density, coming from the matter degrees of freedom
that are trapped on the brane, is given by:
$\rho_T=\rho_\phi+\rho_M$, where
$\rho_\phi=\frac{1}{2}\dot\phi^2+V(\phi)$ is the scalar field
energy density and $V$ is the self-interaction potential, and
$\rho_M=M a^{-3\gamma}$, where $\gamma$ is the barotropic index of
the "ordinary matter" perfect fluid. The Einstein Field Equations
(EFE) for a FRW brane universe are then \cite{maartens}; the
Friedmann Equation:

\begin{equation}
H^2=\frac{k_0}{3}\rho_T(1+\sigma_b
\rho_T)+\frac{\epsilon}{a^4}-\frac{k}{a^2}+\frac{\Lambda}{3},
\label{2.1}
\end{equation}
the Raychaudhuri Equation:

\begin{equation}
2\dot
H=-k_0(\rho_T+p_T)(1+2\sigma_b\rho_T)-\frac{4\epsilon}{a^4}+\frac{2k}{a^2},
\label{2.2}
\end{equation}
and the Klein-Gordon (KG) Equation for the scalar field:

\begin{equation}
\ddot\phi+3H\dot\phi+\frac{dV}{d\phi}=0, \label{2.3}
\end{equation}
where $k_0=8\pi/m_{pl}^2$ ($m_{pl}$ is the effective 4d Planck
mass), $k=0,\pm 1$ is the spatial curvature, $\epsilon$ is a
constant parameter related with black hole mass in the bulk (the
corresponding term in (\ref{2.1}) is known as dark
radiation)\footnotemark\footnotetext{Dark radiation is a
contribution of non-local bulk effects onto the brane, it carries
scalar modes from bulk gravitons. $\epsilon\neq 0$ for
AdS-Schwarzschild bulk, i. e., a AdS bulk with a black hole.} and
$\sigma_b=1/2\lambda_b$ ($\lambda_b$ is the brane tension). The
parameter $\sigma_b$ is related to $k_0$ through the relation
$k_0\sigma_b=\frac{16\pi^2}{3 M_{(5)}^6}$ \cite{brax}. The
pressure of the scalar field fluid is
$p_\phi=\frac{1}{2}\dot\phi^2-V(\phi)$. Note that, if we drop the
dark radiation term, when $\rho_T\ll 1/\sigma_b$ the standard GR
Friedman equation is recovered.

We will look for flat ($k=0$) FRW solutions to the brane EFE's
(\ref{2.1},\ref{2.2},\ref{2.3}) generalizing the method developed
by Chimento and Jakubi in Ref.\cite{chimento}.

\subsection{Method for generating exact solutions}\label{sec2.1}

The idea behind the method in Ref.\cite{chimento} is the
following. If one rewrites the self-interaction potential as a
function of the scale factor $a$: $V(a)=\frac{F(a)}{a^6}$, where
$F(a)$ is an "arbitrary" input function, then, it is not difficult
to prove, that (by working out of the KG equation) the scalar
field energy density can be found as a first integral of (2.3):

\begin{equation}
\rho_\phi=\frac{1}{a^6}\left\{6\int\frac{da}{a}F(a)+C\right\},
\label{2.4}
\end{equation}
where $C$ is an arbitrary integration constant. Let us introduce
the following "Chimento's functions" (CFs):

\begin{equation}
G(a)\equiv
H^2,\;\;L(a)\equiv\dot\phi^2=2\left\{\rho_\phi-\frac{F(a)}{a^6}\right\},\label{2.5}
\end{equation}

Then, using the scale factor as an independent variable, the
problem of finding solutions to the EFE, can be reduced to
quadratures:

\begin{equation}
\triangle t=\int\frac{da}{a}G(a)^{-1/2},\label{2.6}
\end{equation}
and

\begin{equation}
\triangle\phi=\int\frac{da}{a}\left[\frac{L(a)}{G(a)}\right]^{1/2},\label{2.7}
\end{equation}
where $\triangle t\equiv t-t_0$, $\triangle\phi\equiv\phi-\phi_0$,
$t_0$ and $\phi_0$ being two other arbitrary integration
constants. Given an input function $F(a)$, the integration
constant $C$, and the parameters $M$, $\gamma$, $k$, $\epsilon$
and $\Lambda$, one can obtain a solution for the functions $a(t)$,
and $\phi(t)$ and identify the potential $V(\phi)$ as
well.\footnotemark\footnotetext{We want to note here that, unlike
in reference \cite{chimento}, in the present setup (RS2 with AdS
bulk), we were not able to study arbitrary potentials but, more
likely, we were able to identify the self-interaction potential
after applying the method. So, in a sense, the functional form of
the potential is an outcome of the method also.} As we see the
method is fairly general, its only limitation coming from the
choice of the input function $F(a)$ \cite{chimento,tmatos}.

\section{Solutions}\label{soluciones}

We will study a RS2 with AdS bulk. i. e., we will not consider
here the dark radiation term. During inflation this term rapidly
redshifts to zero \cite{lidsey} and, besides, since the decoupling
of matter and radiation, this term could be neglected too. At
early times during evolution $\epsilon$ is not anymore a constant
\cite{sopuerta} and this term does not coincide with that in
Eq.(\ref{2.1}). We set the arbitrary integration constant in
(\ref{2.4}) $C=0$, and the cosmological constant on the brane
$\Lambda =0$. Note that, if cuadratic density contribution could
be neglected (this seems to be the case since nucleosynthesis
epoch), then, a $\Lambda\neq 0$ could be absorbed into the scalar
field self-interaction potential. We consider, also, $k=0$ to
agree with the observational evidence about a spatially flat
universe. Following the original paper \cite{chimento} we set

\begin{equation}
F=B a^s,\label{3.1}
\end{equation}
where $B$ and $s$ are arbitrary constants. By taking explicitly
the integral (\ref{2.4}) one founds: $\rho_\phi=\rho_0\;a^{s-6}$.
It is worth noting that this functional dependence of the scale
factor leads to the following "conservation equation":
$\dot\rho_\phi+(6-s)H \rho_\phi=0$, so the equation of state
$\omega_\phi=\gamma_\phi-1=const\Rightarrow
3\gamma_\phi=6-s=const$. This fact, in turn, leads to the ratio of
the scalar field kinetic energy density and the total scalar field
energy density being a constant:
$\dot\phi^2/2\rho_\phi=\gamma_\phi/2=(6-s)/6=const$.\footnotemark\footnotetext{In
order to study scalar fields with a dynamical equation of state
one should consider the case when the integration constant $C\neq
0$ in equation (\ref{2.4})} In consequence,

\begin{equation}
V(a)=\frac{s}{6}\rho_\phi=\frac{2-\gamma_\phi}{2}\rho_0
a^{-3\gamma_\phi}.\label{3.2}
\end{equation}

Since we do not need to give the functional form of the
self-interaction potential (as a function of $\phi$) as an input,
then, we do not organize the "gallery" of solutions by grouping
under a given kind of potential (as in \cite{chimento}, for
instance). Instead we study separately the cases when 1) both
ordinary matter and the scalar field are sources of the Einstein's
equations on the brane, and 2) only a scalar field fluid is
trapped on the brane. The CFs are:

\begin{equation}
G(a)=\frac{k_0}{3}\rho_T(1+\sigma_b\rho_T),\;\;L(a)=\gamma_\phi
\rho_\phi.\label{3.3}
\end{equation}

In the case when we have ordinary matter and a scalar field fluid
trapped on the brane, integrals (\ref{2.6}) and (\ref{2.7}) with
CFs (\ref{3.3}) can be explicitly found just for few particular
cases where one fixes a relationship among $\gamma_\phi$ and
$\gamma$. For arbitrary relationships among the barotropic
parameters, exact solutions can be found only for the two limiting
situations: a) the "low-energy" limit where $\rho_T\ll
1/\sigma_b$, and b) the "high-energy" limit, where $\rho_T\gg
1/\sigma_b$. For a universe filled with just a scalar field we are
able to derive a single analytic solution holding "every time"
during the cosmic evolution.

\subsection{\bf Universe filled with ordinary matter and a scalar field}
\label{sec3.1}

\bigskip

\subsubsection{\bf Low-energy limit ($\rho_T\ll
1/\sigma_b\;\Rightarrow\;G(a)\approx\frac{k_0}{3}\rho_T$)}\label{3.1.1}

\bigskip

In this case we are neglecting brane effects so, we recover a case
that has been formerly studied in \cite{chimento} and
\cite{tmatos}. The integrals (\ref{2.6}) and (\ref{2.7}) can be
explicitly found \cite{chimento} (recall that $\rho_T=\rho_0
a^{-3\gamma_\phi}+M a^{-3\gamma}$):

\begin{eqnarray}
\triangle t&=&\sqrt{\frac{4}{3k_0
\rho_0}}\frac{a^{3\gamma_\phi/2}}{\gamma_\phi}\times\nonumber\\&_2F_1&
\left(\frac{1}{2},\frac{\gamma_\phi}{2(\gamma_\phi-\gamma)},
\frac{3\gamma_\phi-2\gamma}
{2(\gamma_\phi-\gamma)},-\frac{M}{\rho_0}a^{3(\gamma_\phi-\gamma)}\right),\nonumber\\
\label{3.4}
\end{eqnarray}
where $\;_2F_1$ is the hipergeometric function, and

\begin{equation}
\triangle\phi=\frac{2\sqrt{(6-s)/k_0}}{3(\gamma-\gamma_\phi)}\;
\sinh^{-1}[\sqrt{\rho_0/M}
a^{3(\gamma-\gamma_\phi)/2}].\label{3.5}
\end{equation}

In consequence we can find the explicit form of the potential as
function of $\phi$:

\begin{equation}
V(\phi)=V_0 \sinh^{2q}[\lambda\triangle\phi],\label{3.6}
\end{equation}
where \[V_0\equiv
\left(\frac{2-\gamma_\phi}{2}\right)\left(\frac{M^{\gamma_\phi}}{\rho_0^\gamma}\right)^
{1/(\gamma_\phi-\gamma)},\]\[\lambda\equiv\frac{3(\gamma-\gamma_\phi)}
{2\sqrt{3\gamma_\phi/k_0}}\] and\footnotemark\footnotetext{When
the scalar field dominates the evolution, it is necessary to
change $\gamma_\phi\rightarrow\gamma$ in the solution.}
\[q=\gamma_\phi/(\gamma_\phi-\gamma).\]

As explained in \cite{tmatos}, this potential is a good
quintessencial candidate to be the missing energy in the universe.
Since it behaves like an inverse power-law potential at early
times, then this allows to avoid the fine tuning problem. The
parameters $V_0$, $\lambda$ and $q$, can be determined uniquely by
the measured values for the equation of state and the amount of
vacuum energy to obtain a tracker solution in such a way as to
avoid the coincidence problem as well \cite{tmatos}. A good
agreement with current observations of SNIa, Angular and Mass
power spectrums was reported in reference \cite{tmatos}.

\bigskip

\subsubsection{\bf High-energy limit ($\rho_T\gg
1/\sigma_b\;\Rightarrow\;G(a)\approx\frac{k_0}{3}\sigma_b\rho_T^2$)}\label{3.1.2}

\bigskip

We concentrate in this case since brane effects dominate. Let us
introduce a new variable $X\equiv a^{3\gamma}$ and a constant
parameter $m=1-\frac{\gamma_\phi}{\gamma}$, then the CFs take the
following form:

\begin{eqnarray}
L(X)&=&\gamma_\phi \rho_0 X^{m-1},\nonumber\\ G(X)&=&
(k_0/3)\sigma_b \rho_0^2 X^{-2} (X^m+(M/\rho_0))^2.\label{3.7}
\end{eqnarray}

In terms of the new variable and constant parameter the integrals
(2.6) and (2.7) look like;

\begin{equation}
\triangle
t=\frac{1}{\gamma\rho_0\sqrt{3k_0\sigma_b}}\int\frac{dX}{X^m+(M/\rho_0)},\label{3.8}
\end{equation}
and

\begin{equation}
\triangle\phi=(1/\gamma)\sqrt{\frac{\gamma_\phi}{3k_0\sigma_b\rho_0
}}\int\frac{dX X^\frac{m-1}{2}}{X^m+(M/\rho_0)},\label{3.9}
\end{equation}
respectively. These integrals can now be explicitly taken in the
general case to yield:

\begin{equation}
\triangle t=\frac{a^{3\gamma}}{\sqrt{3k_0\sigma_b}\gamma
M}\;_2F_1\left(1,\frac{\gamma}{\gamma-\gamma_\phi},
\frac{2\gamma-\gamma_\phi}{\gamma-\gamma_\phi},
-\frac{\rho_0}{M}a^{3(\gamma-\gamma_\phi)}\right), \label{3.10}
\end{equation}
and

\begin{eqnarray}
\triangle\phi&=&\pm\frac{(2/M)}{2\gamma-\gamma_\phi}
\sqrt{\frac{\gamma_\phi\rho_0}{3k_0\sigma_b}}a^{3\gamma}\times
\nonumber\\&_2F_1&\left(1,\frac{2\gamma-\gamma_\phi}{2(\gamma-\gamma_\phi)},
\frac{4\gamma-3\gamma_\phi}{2(\gamma-\gamma_\phi)},
-\frac{\rho_0}{M}a^{3(\gamma-\gamma_\phi)}\right).\nonumber\\
\label{3.11}
\end{eqnarray}
where the "+" sign should be taken if $2\gamma>\gamma_\phi$ and
the "-" if the inverse inequality holds. This is a new exact
solution to "high-energy" brane cosmology. Since it is general in
the sense that the form of the self-interaction potential is not
specified apriori and, besides, the method used to derive
solutions is general and free of any assumption (except the
assumed form of the input function $F(a)$), it generalizes former
solutions, for instance, those in \cite{hawkins,lidsey,savchenko}.
Besides, in \cite{hawkins,lidsey}, since the authors were
interested in describing scalar driven inflation, only one (scalar
field) fluid was assumed to be confined to the RS2 brane.

Although, in this case, we are not able to identify a
self-interaction potential of a simple (analytic) form, we can
judge about its asymptotic behavior. In fact, by taking the
asymptotic of the hypergeometric function $\;_2F_1(a,b,c,z)$ for
small $z\propto a^{3(\gamma-\gamma_\phi)}$ in Eq. (\ref{3.11}), in
the first approximation,\footnotemark\footnotetext{We are
considering that the scale factor $a$ is normalized in such a way
that its present value $a_0=1$. In consequence, taking small $z$,
i. e., small $a\ll 1$, is consistent with the high-energy limit we
are considering in this subsection.} one obtains that
$\triangle\phi\propto a^{3\gamma}$. In consequence, at early times
during the evolution, the behavior of the self-interaction
potential is like $V(\phi)\propto
(\triangle\phi)^{-\gamma_\phi/\gamma}$. This kind of potentials
has been studied in \cite{lidsey}. It has been established therein
that, inflation is possible for $\gamma_\phi>2\gamma$, if
\[\phi<\phi_e\approx \left(\frac{k_0\sigma_b
\gamma^2}{\gamma_\phi^2}V_0\right)^{\frac{\gamma}{\gamma_\phi-2\gamma}},\]
where
\[V_0\equiv\frac{(\gamma_\phi-2\gamma)M}{2}\sqrt\frac{3k_0\sigma_b}{\gamma_\phi\rho_0}.\]
It follows then, that if the inverse of the brane tension
$\sigma_b$;
\[\sigma_b<\frac{2}{V_0}\left(\frac{\gamma_\phi}{\gamma\sqrt{2k_0}}\right)^
{\gamma_\phi/\gamma},\] the universe inflates for
$\phi<\phi_e$.\footnotemark\footnotetext{Here we are considering
the results of \cite{lidsey} but with appropriated identification
of the constants with the ones in the present paper.} The
observational constraints on a model of brane inflation based upon
this kind of potential are discussed in detail in reference
\cite{lidsey}. However, in that reference the authors impose the
constraints supposing that this potential can take account of both
early inflation and late time inflation. We should note that, in
the case of interest here, this kind of potential emerges just an
asymptotic form of a more general potential an it can take account
only of the early inflation so, some of the constraints imposed in
reference \cite{lidsey}, might not apply in this case so we get a
wider range for the parameters. This point will be discussed in
some detail in the next section.

In order to have simpler expressions and to obtain simple explicit
functional dependence of $V$ on $\phi$, we can explore particular
values of $m$ in (\ref{3.7},\ref{3.8},\ref{3.9}) (equivalent to
taking fixed relationships among the barotropic parameters), i)
When $\gamma_\phi=\gamma\;\Rightarrow m=0$ in (\ref{3.7}), so the
integrals (\ref{3.8}) and (\ref{3.9}) yield:

\begin{equation}
\triangle t=\frac{a^{3\gamma}}{\sqrt{3k_0\sigma_b}\gamma A},
\label{3.12}
\end{equation}
and

\begin{equation}
\triangle\phi=\sqrt\frac{4\rho_0}{3k_0\gamma\sigma_b}\frac{a^{3\gamma/2}}{A}.
\label{3.13}
\end{equation}

The self-interaction potential is a power law one:

\begin{equation}
V(\phi)=V_0 (\triangle\phi)^{-2},\;\;V_0\equiv\frac{2(2-\gamma)}
{3k_0\sigma_b\gamma}\frac{\rho_0^2}{A^2}, \label{3.14}
\end{equation}
where, as before, $A=M+\rho_0$.

The case with $m=1$ ($\gamma_\phi=0$) is trivial. In this case we
have a constant potential $V(\phi)=\rho_0=const$. For the time
evolution of the scale factor, we have, according to (\ref{3.8}):

\begin{equation}
\triangle
t=\frac{\ln[a^{3\gamma}+M/\rho_0]}{\sqrt{3k_0\sigma_b}\gamma\rho_0}.
\label{3.15}
\end{equation}

\bigskip

For other fixed relationships among the $\gamma$s one has;

\bigskip

ii) $m=-1$ ($\gamma_\phi=2\gamma$).

\bigskip

In this case:

\begin{equation}
\triangle t=\frac{1}{\sqrt{3k_0\sigma_b}\gamma
M}\left(Ma^{3\gamma}-\frac{\rho_0}{M}
\ln[a^{3\gamma}+\rho_0/M]\right), \label{3.16}
\end{equation}
and

\begin{equation}
\triangle\phi=\frac{1}{M}\sqrt\frac{2
\rho_0}{3k_0\gamma\sigma_b}\ln[a^{3\gamma}+\rho_0/M]. \label{3.17}
\end{equation}

The self-interaction potential is of the following form:

\begin{equation}
V(\phi)=\frac{V_0}{(\exp[\lambda\triangle\phi]-b)^2}, \label{3.18}
\end{equation}
where $V_0\equiv (1-\gamma)\rho_0$, $\lambda\equiv
M\sqrt{\frac{3k_0\sigma_b \gamma}{2\rho_0}}$ and the constant
$b\equiv \rho_0/M$. Its asymptotic behavior is like in subsection
\ref{3.1.2} (see discussion under equation (\ref{3.11})) for
$\lambda\triangle\phi\ll 1$, $\gamma_\phi=2\gamma$, and like in
(\ref{3.6}) for $\lambda\triangle\phi\gg 1$, $q=-1$. Therefore,
this potential falls into the category of "unified" potentials
that will be discussed below.

\bigskip

iii) $m=-\infty$ ($\gamma=0$, $\gamma_\phi$ arbitrary)

\bigskip

In this particular case the background fluid is a vacuum fluid. In
order to avoid using the formal limit $m=-\infty$, it is
convenient to change variables $X\rightarrow Y=X^{1-m}$ and
$m\rightarrow n=\frac{m}{m-1}$. The integrals (\ref{2.6}) and
(\ref{2.7}) ((\ref{3.8}) and (\ref{3.9})) can be cast now, into
the following form:
\[\triangle t=\frac{1}{\sqrt{3k_0\sigma_b}M
\gamma_\phi}\int\frac{dY}{Y^n+(\rho_0/M)} \] and
\[\triangle\phi=\frac{1}{M}\sqrt{\frac{\rho_0}{3k_0\sigma_b
\gamma_\phi}}\int\frac{dY}{\sqrt{Y}(Y^n+(\rho_0/M))}\]
respectively. After integration for $n=1$ ($m=-\infty$);

\begin{equation}
\triangle
t=\frac{\ln[a^{3\gamma_\phi}+\rho_0/M]}{\sqrt{3k_0\sigma_b}\gamma_\phi
M}, \label{3.19}
\end{equation}

\begin{equation}
\triangle\phi=\left(\frac{2}{\sqrt{3k_0\sigma_b\gamma_\phi
M}}\right)\arctan\left[\sqrt{\frac{M}{\rho_0}}a^{3\gamma_\phi/2}\right],
\label{3.20}
\end{equation}
and, the self-interaction potential is of the following form:

\begin{eqnarray*}
V(\phi)&=&V_0\tan^{-2}[\lambda\triangle\phi]\\ V_0&\equiv&
\left(\frac{2-\gamma_\phi}{2}\right)M,\\
\lambda&\equiv&\frac{\sqrt{3k_0\sigma_b\gamma_\phi M}}{2}.
\label{3.21}
\end{eqnarray*}

This kind of potential has not been formerly studied under the
brane perspective, but we leave it for further research.

\subsection{Scalar Field Dominated Solution}\label{sec3.3}

\bigskip

Let us now consider the case when the scalar field dominates the
matter content. We recall that, since we set to zero the
integration constant in Eq. (\ref{2.4}), then the equation of
state of the scalar field is a constant (meaning constant ratio of
kinetic to total energy density of the scalar field). Let us
introduce a new variable $Y\equiv a^{3\gamma_\phi}$, then the CFs
are:

\begin{eqnarray}
G(Y)&=&A(Y+B)/Y^2,\nonumber\\L(Y)&=&C Y^{-1}, \label{3.25}
\end{eqnarray}
where the constants $A\equiv\frac{k_0\rho_0}{3}$,
$B\equiv\sigma_b\rho_0$, and $C\equiv\gamma_\phi\rho_0$.
Meanwhile, in terms of the new variable $Y$, the integrals (2.6)
and (2.7) can be written as follows:

\begin{eqnarray}
\triangle t&=&\frac{1}{3\gamma_\phi}\int\frac{dY}{Y\sqrt{G(Y)}},\nonumber\\
\triangle\phi&=&\frac{1}{3\gamma_\phi}
\int\frac{dY}{Y}\sqrt\frac{L(Y)}{G(Y)}. \label{3.26}
\end{eqnarray}

These integrals can be taken exactly to yield:

\begin{equation}
\triangle t=\frac{2}{3\gamma_\phi\sqrt{A}}\sqrt{Y+B}, \label{3.27}
\end{equation}
and

\begin{eqnarray}
\triangle\phi&=&\frac{2}{3\gamma_\phi}
\sqrt\frac{C}{A}\ln\left[\sqrt{Y}+\sqrt{Y+B}\right]\nonumber\\&=&
\frac{2}{3\gamma_\phi} \sqrt\frac{C}{A}
\sinh^{-1}\left[\sqrt\frac{Y}{B}\right]. \label{3.28}
\end{eqnarray}

If one realizes that the self-interaction potential can be written
as $V(Y)=V_0 Y^{-1}$, where
$V_0\equiv\frac{2-\gamma_\phi}{2\gamma_\phi}\;C$, then, the last
equation can be rewritten as:

\begin{equation}
\frac{V}{\left(1+\sqrt{1+\frac{2\sigma_b}{2-\gamma_\phi}V}\right)^2}=V_0
e^{-2\lambda\triangle\phi}, \label{3.29}
\end{equation}
where, $\lambda\equiv
3\gamma_\phi\sqrt\frac{A}{C}=\sqrt{3\gamma_\phi k_0}$. Equation
(\ref{3.28}) can be rewritten also in a form where the functional
form of the self-interaction potential is straightforward:

\begin{eqnarray}
V(\phi)&=&\bar V_0 \sinh[\lambda\triangle\phi]^{-2},\nonumber\\
\bar V_0&\equiv&\frac{2-\gamma_\phi}{2\sigma_b}. \label{3.30}
\end{eqnarray}

This solution has been formerly obtained in
\cite{hawkins}.\footnotemark\footnotetext{An appropriated
identification of the constants should be done. The constant that
appears in \cite{hawkins} $C^2=3\gamma_\phi$. Do not confound with
the constant $C$ in the present paper.} It is interesting because
it reduces to the power-law cosmology driven by an exponential
potential at late times (in the low-energy limit), meanwhile, at
early times the asymptotic behavior is like $a\sim
t^{1/3\gamma_\phi}$, and inflation proceeds for a finite time for
$\gamma_\phi<1/3$ \cite{hawkins}. In this limit (high-energy
limit) the self-interaction potential $V\propto\phi^{-2}$. Models
with the same asymptotics has been studied in \cite{kdimopoulos},
to account for a unified description of early inflation and late
time accelerated expansion.

However, this simple model of the cosmic evolution can not really
account for a unified description of early inflation and
quintessential inflation, since a scalar field with a non-evolving
equation of state can not describe correctly a period of cosmic
evolution starting in a inflaton dominated regime, followed by a
matter dominated regime (including a period of radiation
domination) and leading to a cosmological constant dominated era.
To have a scalar field with an evolving equation of state one
should consider the integration constant in (\ref{2.4}) different
from zero or, instead, to study other kinds of input function
$F(a)$.

\section{Unified Description of Inflation and Quintessence}\label{unifi}

\bigskip

The main point we want to discuss here is the "unified" picture
that comprises both early inflation and late time (quintessential)
acceleration of the expansion. This idea has been clearly stated
in \cite{peebles} for standard "non-brane" models and, in brane
contexts, for instance, in references
\cite{lidsey,sss,kdimopoulos}. The fact is that, the solutions we
found with the help of the general method of reference
\cite{chimento}, share the same asymptotics of the models studied
in \cite{hawkins,lidsey,sss,kdimopoulos}, that lead to such a
unified approach to inflation. The most general solutions we found
in the present paper were the limiting "low-energy", and
"high-energy" solutions given in subsections \ref{3.1.1} and
\ref{3.1.2}, respectively so we will concentrate our discussion in
this case.\footnotemark\footnotetext{By "general" we understand
here a solution to the field equations with two sources: ordinary
fluid and scalar field fluid, that does not depend upon the
relationship among the barotropic parameters of the background and
scalar field fluids.} It is obvious that, to describe the
early-time evolution, it is the "high-energy" solution the
adequate one. Although a simple behavior of the solution is not at
hand, the asymptotic behavior for $a\ll 1$ leads to
$V\propto(\triangle\phi)^{-\gamma_\phi/\gamma}$. This is the kind
of asymptotics that is able to account for the stage of early
inflation by appropriately choosing the free parameters, in order
to agree with the observational constraints \cite{hawkins}. We can
write this potential as follows

\begin{eqnarray}
  V(\phi)&=&\frac{M^{4+n}}{\phi^n},\nonumber\\n&\equiv&\frac{\gamma_\phi}{\gamma},
  \nonumber\\M^{4+n}&\equiv& \left(\frac{2-\gamma_\phi}{2}\right)\left[\frac{(2/M)}{|2/n-1|}
  \frac{1}{\sqrt{3k_0\sigma_b\gamma_\phi}}\right]^n\rho_0^{1+n/2}\nonumber\\.
\end{eqnarray}

In standard inflation chaotic scenario based on GR, inflation
proceeds when $\alpha^2\equiv (1/k_0)(V'/V)^2\ll 2$ and ends when
$\alpha^2\approx 2$. However, in cosmological FRW brane models of
RS2 type, one can introduce an effective slow-roll parameter
\cite{lidsey}:

\begin{equation}
\bar\alpha^2\equiv\frac{\alpha^2}{1+\sigma_b\rho}.
\end{equation}

Hence, if $\sigma_b\rho\gg 1$, there may be inflation even though
$\alpha^2\gg 2$. The end of inflation is characterized by the
condition $\bar\alpha^2\approx 1$, which in the case under study
($\alpha^2=(n^2/k_0)(1/\phi^2)$), yields that the scalar field
value at the end of inflation $\phi_f^{n-2}\approx k_0\sigma_b
M^{4+n}/n^2$. The number of e-foldings between a scalar field
value $\phi$ and $\phi_f$: $N\approx
-k_0\sigma_b\int_\phi^{\phi_f}d\phi V^2/V'$ can be written in the
form of a relationship between $\phi$ and $\phi_f$:

\begin{equation}
\left(\frac{\phi}{\phi_f}\right)=\frac{1}{[1+(\frac{n-2}{n})N]^{1/(n-2)}}.
\end{equation}

An observationally acceptable lower limit of the scalar spectral
index ($n_s\approx 1-\frac{2(2n-1)/n}{1+(\frac{n-2}{n})N}$) is
$n_s\approx 0.9$. This bound on $n_s$ translates into a lower
bound on $n$ given a fixed $N$. Observations constraint the
spectral tilt at $N\approx 50$. In this case $n>7$ \cite{lidsey}.
For bigger $N$s this lower limit is smaller. For instance, for
$N=69$, $n>3$. Another constraint imposed to the model is given by
the following condition: $V(\phi) <
m_5^4=(\frac{32\pi^2}{6\sigma_bk_0})^{2/3}$. Otherwise, the
assumption that the inflaton field is confined to the brane may
not be valid above the five-dimensional Plack scale \cite{lidsey}.
The magnitude of the self-interaction potential $N=50$ e-foldings
before the end of inflation:

\begin{equation}
V_N\approx 8\left[\left(\frac{600
\pi^2}{4k_0^4\sigma_b^2}\right)\frac{n}{2+(n-2)N}\right]^{1/3}A_{s,COBE}^{2/3},
\end{equation}
where $A_{s,COBE}=2\times 10^{-5}$ is the value of the amplitude
of scalar perturbations after constraints due to COBE
normalization are considered. Thus, the smallness of the
perturbations on COBE scales, ensures that the aforementioned
condition imposed to the self-interaction potential ($V < m_5^4$),
is automatically satisfied and sufficient inflation is possible
\cite{lidsey}. Up to this moment we have discussed about the
"high-energy" limit of the picture, which is suitable to describe
early inflation.

On the opposite hand, to describe the late-time stage of the
cosmic evolution, the most adequate is the "low-energy" solution
of subsection \ref{3.1.1} that, as it has been discussed in
\cite{tmatos}, is a good quintessential candidate to be the
missing energy in the universe. In this case one chooses the
appropriate values for the parameters of the potential in such a
way that the solution of \ref{3.1.1} can only be reached until the
matter and the scalar field energies are of the same order, as it
happens at present. Thus all parameters can be determined and the
solution becomes a tracker one \cite{tmatos}. The behavior of the
scalar field for radiation and matter dominated epochs is the same
than that found for an inverse power-law potential in
\cite{steinhardt}. This "low-energy" model of late time
accelerated inflation was found to be in good agreement with
observational evidence for SNIa, Angular and Mass power spectrums
\cite{tmatos}.

This is the way in which a unified picture of cosmic inflation
emerges with the help of the general method of generating FRW
solutions developed in reference \cite{chimento}, when applied to
RS2 brane cosmologies (with a AdS bulk geometry). Besides, as in
other papers on RS2 cosmological models (for instance
\cite{hawkins}), we have not considered the so called slow-roll
approximation, or any other approximation, to describe the
early-time inflation epoch. It follows then that such unified
picture of cosmic inflation could be a generic property of brane
inflation.

\section{Conclusions}\label{conclusiones}

We have used a fairly general method to derive exact cosmological
FRW solutions in RS2 brane models with a AdS bulk geometry. The
only assumption is given by the choice of the input function
$F(a)$. The method was developed in \cite{chimento}, where it was
formerly used to generate solutions only in standard GR.

We found new exact solutions that generalize formerly known
solutions \cite{hawkins,lidsey,sss,kdimopoulos,savchenko} and were
able to recover other well known solutions like in
\cite{chimento,hawkins,tmatos}. The most interesting fact is that
the solutions found can accommodate both the inflaton field and
quintessence in a common framework (see discussion in the former
section). Since, as stated above, the method used to generate
cosmological solutions is quite general, it seems that such a
unified description of inflation and quintessence is a generic
property of brane inflation cosmological models.

As it was noted in the original paper \cite{chimento}, and has
been discussed, for instance in \cite{hawkins}, there are clear
the limitations imposed by the slow-roll approximation, and most
of the solutions we found cannot be obtained by using this
approximation. In this sense, the models that emerge, and that are
able to describe early-time inflation, are more general than those
obtained within the framework of slow-roll inflation.

The method we have used here could be applied to other brane
cosmological models and, also, other general input functions
(other than $F(a)=Ba^s$ in (\ref{3.1})) could be considered, in
order to obtain other solutions (and, correspondingly, other
self-interaction potentials) than those obtained in the present
paper. Precisely, one of the drawbacks of the present approach is
that the most interesting solution; that containing both scalar
field and ordinary matter as sources, is not a single solution
accounting for the complete cosmic evolution, but consists of two
separated pieces: high-energy and low-energy asymptotic solutions
respectively. The matching of both pieces of the cosmic evolution
is not clear. In this sense, the study of scalar fields with a
dynamical equation of state is of relevance and this is possible
only under other choices of the input function than the one taken
here, or by relaxing the choice of a zero value for the
integration constant in the first integral of the scalar field
energy density.

TM and IQ acknowledge partial financial support by CONACYT
M\'exico, under grants 32138-E and 42748. TM also thanks the
support of the Germany-M\'exico bilateral project DFG-CONACYT 444
MEX-13/17/0-1. IQ and AG acknowledge the MES of Cuba for partial
financial support of this research.





\begin{thebibliography}{99}


\bibitem{wmap} C L Bennett et al, Astrophys. J. Suppl. {\bf 148} (2003)
1 (astro-ph/0302207); G Hinshaw et al, Astrophys. J. Suppl. {\bf
148}(2003)135 (astro-ph/0302217); D N Spergel et al, Astrophys. J.
Suppl. {\bf 148}(2003)175 (astro-ph/0302209); H V Peiris et al,
Astrophys. J. Suppl. {\bf 148}(2003)213 (astro-ph/0302225); A
Kogut et al, Astrophys. J. Suppl. {\bf 148}(2003)161
(astro-ph/0302213); E Komatsu et al, Astrophys. J. Suppl. {\bf
148}(2003)119 (astro-ph/0302223).

\bibitem{inflation} A A Starobinsky, Phys. Lett. B{\bf
91}(1980)99; A H Guth, Phys. Rev. D{\bf 23}(1981)347; A Albrecht
and P J Steinhardt, Phys. Rev. Lett. {\bf 48}(1982)1220; A D
Linde, Phys. Lett. B{\bf 108}(1982)389, {\it ibid} B{\bf
129}(1983)177.

\bibitem{randall} L Randall and R Sundrum, Phys. Rev. Lett. {\bf
83}(1999)3370 (hep-ph/9905221).

\bibitem{hawkins} R M Hawkins and J E Lidsey, Phys. Rev. D{\bf
63}(2001)041301 (gr-qc/0011060).

\bibitem{lidsey} G Huey and J E Lidsey, Phys. Lett. B{\bf 514}(2001)217-225
(astro-ph/0104006).

\bibitem{ford} L H Ford, Phys. Rev. D{\bf 35}(1987)2955.

\bibitem{feng} B Feng and M Li, Phys. Lett. B{\bf
564}(2003)169-174 (hep-ph/0212213).

\bibitem{urena} A R Liddle and L A Urena-Lopez, Phys. Rev. D{\bf
68}(2003)043517 (astro-ph/0302054).

\bibitem{sami} M Sami and V Sahni, hep-th/0402086.

\bibitem{majumdar} A S Majumdar, Phys. Rev. D{\bf 64}(2001)083503
(astro-ph/0105518).

\bibitem{sss} V Sahni, M Sami and T Souradeep, Phys. Rev. D{\bf
65}(2002)023518 (gr-qc/0105121).

\bibitem{kdimopoulos} K Dimopoulos, astro-ph/0210374;
Phys. Rev. D{\bf 68}(2003)123506 (astro-ph/0212264);K Dimopoulos
and J W F Valle, Astropart. Phys. {\bf 18}(2002)287-306
(astro-ph/0111417).

\bibitem{peebles} P J E Peebles and A Vilenkin, Phys. Rev. D{\bf 59}
(1999)063505 (astro-ph/9810509).

\bibitem{chimento} L P Chimento and A S Jakubi, Int. J. Mod. Phys.
D{\bf 5}(1996)71-84 (gr-qc/9506015).

\bibitem{tmatos} L A Urena-Lopez and T Matos,  Phys. Rev. D{\bf 62}(2000)081302
(astro-ph/0003364).

\bibitem{savchenko} N Yu Savchenko and A V Toporensky, Class.
Quant. Grav. {\bf 20}(2003)2553-2562 (gr-qc/0212104).

\bibitem{kmaeda} K Maeda, Phys. Rev. D {\bf 64}(2001)123525
(astro-ph/0012313).

\bibitem{maartens} R Maartens, gr-qc/0312059.

\bibitem{brax} P Brax and C van de Bruck, Class. Quant. Grav.
{\bf 20}(2003)R201-R232 (hep-th/0303095).

\bibitem{sopuerta} E Leeper, R Maartens and C F Sopuerta,
Class. Quant. Grav. {\bf 21}(2004)1125-1134 (gr-qc/0309080).

\bibitem{steinhardt} I Zlatev, L Wang and P J Steinhardt, Phys. Rev. Lett.
{\bf 82}(1999)896-899 (astro-ph/9807002); P J Steinhardt, L Wang
and I Zlatev, Phys. Rev. D{\bf 59}(1999)123504 (astro-ph/9812313).

\bibitem{liddle} A R Liddle and D H Lyth, Phys. Rep. {\bf 231}, 1
(1993). R Maartens, D Wands, B A Bassett and I P C Heard, Phys.
Rev. D {\bf 62} 041301 (2000).

\end{thebibliography}
\end{document}